\def\draftversion{false}
\RequirePackage{ifthen}
\ifthenelse{\equal{\draftversion}{true}}{
  \documentclass[aps,prl,galley,superscriptaddress,
       longbibliography]{revtex4-1}
}{
  \documentclass[aps,showpacs,prl,reprint,superscriptaddress,longbibliography]{revtex4-1}

}

\usepackage[colorlinks=true,linkcolor=blue,citecolor=blue,urlcolor=blue]{hyperref}

\usepackage{hyperref}
\usepackage{setspace} %to set 1.5 spacing in text
\usepackage{graphicx}
\usepackage{amsmath}
\usepackage{color}
\usepackage{amsmath}
\usepackage{amssymb}
\usepackage{verbatim}
\usepackage{latexsym}
\usepackage{enumerate} % to change the numbering of 'enumerate'
\usepackage{bm} % for bold face mathematical symbols
\usepackage[caption=false]{subfig}
\usepackage[export]{adjustbox}

\usepackage{soul}  % \st    for strike-out

\ifthenelse{\equal{\draftversion}{true}}{
  \marginparwidth 2.7in
  \marginparsep 0.5in
  \newcounter{comm} % counter for commentaries
  \def\commnext{\stepcounter{comm}}
  \def\commtext{{\bf\color{blue}[\arabic{comm}]}}
  \def\commmar{{\bf\color{blue}[\arabic{comm}]}}
  \def\dvm#1{\commnext\marginpar{\small DV\commmar: #1}\commtext}
  \def\tmm#1{\commnext\marginpar{\small TM\commmar: #1}\commtext}
  \def\mlab#1{\marginpar{\small\bf #1}}
  
}{
  \def\dvm#1{}
  \def\tmm#1{}
  \def\mlab#1{}
  
}
\begin{document}

\title{Phonon-assisted spin splitting in centrosymmetric crystals}

\author{Bartomeu Monserrat} \email{bm418@cam.ac.uk}
\affiliation{Department of Physics and Astronomy, Rutgers University,
  Piscataway, New Jersey 08854-8019, USA} 
\affiliation{TCM Group,
  Cavendish Laboratory, University of Cambridge, J.\ J.\ Thomson
  Avenue, Cambridge CB3 0HE, United Kingdom}
\author{David Vanderbilt} 
\affiliation{Department of Physics and Astronomy, Rutgers University,
  Piscataway, New Jersey 08854-8019, USA} 

\date{\today}

\begin{abstract} 
For static crystals it is well known that
electronic states are doubly degenerate in their spin degree of freedom
in the presence of time reversal and inversion symmetries.
This degeneracy can only be lifted by either (i) breaking time
reversal symmetry, for example in a ferromagnet, or (ii) breaking
inversion symmetry and having spin-orbit coupling, for
example in the Rashba effect. We propose that spin degeneracy can be
lifted in time reversal and inversion symmetric crystals with a
combination of lattice vibrations and spin-orbit coupling. We
demonstrate this effect in the cubic perovskite CsPbCl$_3$ by
performing first principles calculations of the finite temperature
band structure, which, in accordance with our prediction, undergoes
spin splitting. We also suggest optical and photoemission
experiments to examine our predictions. 
%, and relate our findings
%to the recent observation of low electron-hole recombination rates
%in the halide perovskites.
%rationalize recent experiments on the electron-hole recombination rate in lead halide perovskites. 
This new understanding dramatically expands the range of materials
that can exhibit spin splitting, with potential applications in
a variety of technologies such as spintronics and photovoltaics.
\end{abstract}

\maketitle

The microscopic properties of crystals are strongly constrained by
their underlying symmetries. One example concerns the
allowed energy levels for electrons, which are intimately related to
time reversal and inversion symmetries. In the presence of time
reversal symmetry, electrons must obey the Kramers degeneracy
theorem, which implies that electronic energy levels are related by
$\varepsilon_n(-\mathbf{k},-\sigma)=\varepsilon_n(\mathbf{k},\sigma)$.
In this expression, the electron quantum numbers are the band index
$n$, the crystal momentum $\mathbf{k}$, and the spin label
$\sigma$.
%
%\dvm{In general with SOC $\sigma$ is a label for states that have
%spin expectation values up and down but are not spin eigenstates.
%Reword somehow?}
%
In the presence of inversion symmetry, electronic energies
are related by
$\varepsilon_n(-\mathbf{k},\sigma)=\varepsilon_n(\mathbf{k},\sigma)$.
As a consequence, time reversal and inversion symmetric crystals
exhibit spin degenerate electronic states
\begin{equation}
\varepsilon_n(\mathbf{k},-\sigma)=\varepsilon_n(\mathbf{k},\sigma). \label{eq:spin_deg}
\end{equation}

It is commonly held that the lifting of spin degeneracy can be accomplished in two ways. The first is to break time reversal symmetry, for example with a magnetic field $\bm{B}$ which couples to the electron spin with a Zeeman term $-\bm{\mu}_s\bm{\cdot B}$, where $\bm{\mu}_s$ is the electron magnetic moment arising from its spin. This term has opposite signs for opposite spin states, and therefore lifts the spin degeneracy of electronic bands. The second is to break inversion symmetry and include the spin-orbit interaction $-\lambda\bm{\sigma\cdot p\times\mathcal{E}}$~\cite{winkler_book}, where $\lambda=e\hbar/4m^2c^2$ is a fundamental constant, $\bm{\sigma}$ is a vector of Pauli matrices, $\bm{p}$ is the momentum operator, and $\bm{\mathcal{E}}$ is an electric field. The spin-orbit interaction obeys time reversal and inversion symmetries, and therefore by itself cannot break spin degeneracy. The additional breaking of inversion symmetry enables the spin splitting of electron bands, as observed for example in the Rashba or Dresselhaus effects~\cite{rashba_original_paper,rashba_second_paper,dresselhaus_soc,zunger_rashba}.

The central insight of this work is that, even in the presence of time reversal and inversion symmetries, the electronic spin degeneracy can be lifted with a combination of lattice vibrations and spin-orbit coupling. This leads to a \textit{dynamical} spin splitting that obeys Eq.~(\ref{eq:spin_deg}) only on average. 

We consider a two-band model to introduce phonon-assisted spin splitting in centrosymmetric crystals. We work within the Born-Oppenheimer approximation, which implies that the adiabatic principle holds and electrons instantaneously follow the slower nuclear motion. Within this paradigm, we consider a generic nuclear configuration characterized by a set of three Brillouin zone (BZ) center optical phonon amplitudes $(u_1,u_2,u_3)$, which contribute to a spin-orbit coupling term
\begin{equation}
\mathcal{H}_{\mathrm{SOC}}=\lambda_{\mathrm{SOC}}\sum_{ijk}\epsilon_{ijk}u_ik_j\sigma_k.
\end{equation}
In this equation, $\lambda_{\mathrm{SOC}}$ describes the spin-orbit
interaction strength, $\mathbf{k}$ is the electron wave vector
measured from the BZ center, $\bm{\sigma}$ are
the Pauli matrices, and $\epsilon_{ijk}$ is the antisymmetric
Levi-Civita symbol. Focusing on a ray in $\mathbf{k}$-space, 
say along $k_3$, the Hamiltonian reduces to
\begin{equation}
\mathcal{H}_{\mathrm{SOC}}=\lambda_{\mathrm{SOC}}k_3(u_1\sigma_2-u_2\sigma_1).
\end{equation}
This is a Rashba-like Hamiltonian in which the Rashba parameter is $\lambda_{\mathrm{SOC}}k_3$. The full two-band Hamiltonian can be constructed by adding the kinetic energy, leading to a dispersion along $k_3$ given by
\begin{equation}
\varepsilon_{\pm}(k_3)=\frac{\hbar^2k_3^2}{2m}\pm\lambda_{\mathrm{SOC}}k_3\sqrt{u_1^2+u_2^2}. \label{eq:rashba_dispersion}
\end{equation}
The second term determines the amount of spin splitting of the otherwise degenerate bands. We can see that the spin splitting vanishes if there is no spin-orbit coupling ($\lambda_{\mathrm{SOC}}=0$) or if there are no phonons in the system ($u_1=u_2=0)$, demonstrating that both ingredients are necessary. We also see that for the quadratic free electron dispersion, the spin splitting increases with increasing $k_3$. In general, the first term in Eq.~(\ref{eq:rashba_dispersion}) also depends on the nuclear configuration, but as it does not lead to a spin splitting of bands, we ignore this dependency in our model.

The next step is to consider the quantum and thermal average over the phonon amplitudes $(u_1,u_2)$. We assume that the system is harmonic with each normal mode $u$ determining the nuclear density 
\begin{equation}
|\phi(u;T)|^2=\frac{1}{\sqrt{2\pi s^2(T)}}e^{-u^2/2s^2(T)},
\end{equation}
where $s^2(T)=(\hbar/2M\omega)\coth(\hbar\omega/2k_{\mathrm{B}}T)$ is the Gaussian width, $M$ is the reduced mass, and $\omega$ is the vibrational harmonic frequency associated with mode $u$. 
%We note that we use mass-reduced normal mode coordinates $u$. 
For simplicity, we assume that the two modes $(u_1,u_2)$ have the same frequency $\omega$. Therefore, as $(u_1,u_2)$ are Gaussian distributed, the band energies in Eq.~(\ref{eq:rashba_dispersion}) follow the Rayleigh distribution
\begin{equation}
n(\varepsilon)=\frac{|\varepsilon-\varepsilon_0|}{\mu^2}e^{-(\varepsilon-\varepsilon_0)^2/2\mu^2}, \label{eq:rayleigh}
\end{equation}
where $\varepsilon_0=\hbar^2k_3^2/2m$ and $\mu^2=s^2(T)\lambda_{\mathrm{SOC}}^2k_3^2$.

\begin{figure}
\centering
\subfloat[][]{
\includegraphics[width=0.23\textwidth,valign=t]{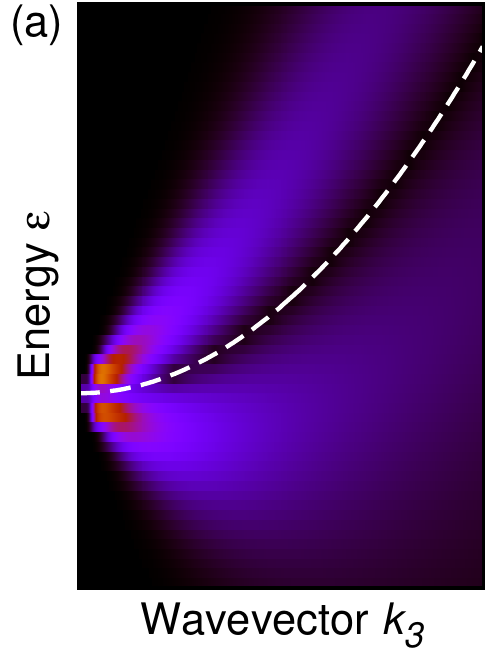}
\label{subfig:dispersion}} 
\subfloat[][]{
\includegraphics[width=0.23\textwidth,valign=t]{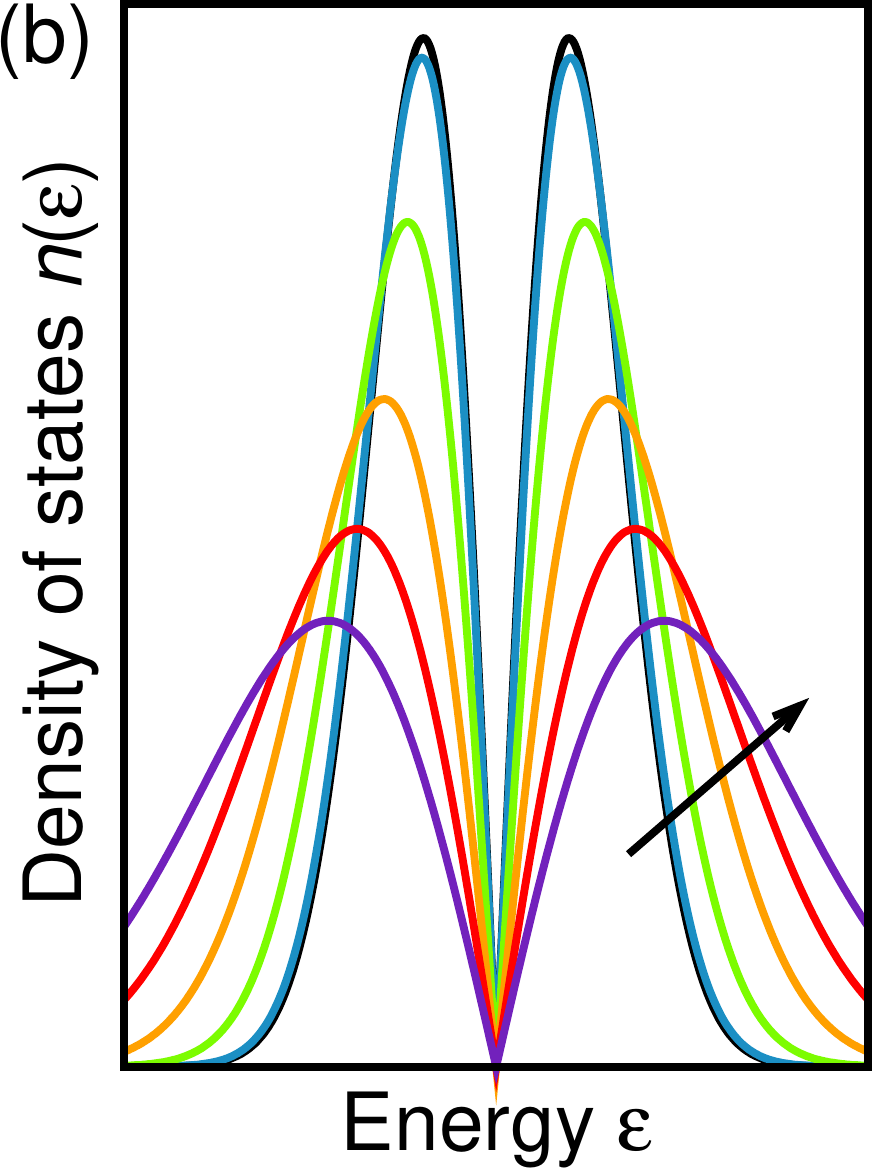}
\label{subfig:tdep}} \\
\caption{(a) Energy dispersion of the two-band model assuming that
the nuclei are static (white dashed line) and calculating the
thermal average in the harmonic approximation (color gradient).
(b) Density of states $n(\varepsilon)$ at fixed $k_3$ of the two-band model as a function of temperature. The black line corresponds to $T=0$\,K (including quantum nuclear effects only), and the other lines correspond to increasing temperatures in steps of $\hbar\omega/2k_{\mathrm{B}}=50$\,K for $\hbar\omega=0.02$\,eV in the direction indicated by the arrow. We note that the expected energy variations of the spin-split bands with temperature will be of the order of $0.1$\,eV in the range $0$-$300$\,K~\cite{bitex_rashba_tdep}.}
\label{fig:2band}
\end{figure}

%\begin{figure}
%\centering
%\includegraphics[scale=0.4]{./2band-tdep.pdf}
%\caption{Density of states $n(\varepsilon)$ at fixed $k_3$ of the two-band model as a function of temperature. The black line corresponds to $T=0$\,K, and the other lines correspond to increasing temperatures in the direction indicated by the arrow.} 
%\label{fig:tdep}
%\end{figure}

The vibrational distribution of the two-band model is depicted in Fig.~\ref{subfig:dispersion}, where the static-lattice doubly degenerate band (white dashed line) is spin-split into two bands when nuclear vibrations are included (color gradient). By analyzing Eq.~(\ref{eq:rayleigh}) and Fig.~\ref{fig:2band} we observe a few properties of the proposed phonon-assisted spin splitting. First, the two bands in Fig.~\ref{fig:2band} are fundamentally distinct from the standard Rashba spin splitting, because there is no definite spin state associated with each. Instead, as phonon amplitudes $u$ and $-u$ have equal weight within the harmonic approximation, and they lead to opposite signs for the linear term in Eq.~(\ref{eq:rashba_dispersion}), the two bands have contributions from both spin states. Second, the minimum band gap within the static-lattice approximation becomes a crossing point for the spin-split bands, whose minima are shifted in both momentum and energy. This is similar to the Rashba effect, and has important consequences for the optical properties associated with these bands. Third, the band splitting is already present at zero temperature, where it is induced by zero-point quantum motion, and grows with increasing temperature. This is depicted in Fig.~\ref{subfig:tdep} where the temperature dependence of the density of states 
is shown at a fixed $k_3$.

We next illustrate phonon-assisted spin splitting in centrosymmetric crystals by performing first principles calculations in the cubic perovksite CsPbCl$_3$. The calculations have been performed using density functional theory (DFT)~\cite{PhysRev.136.B864,PhysRev.140.A1133,dft_rev_mod_phys} and the projector augmented-wave method~\cite{paw_original} as implemented in the {\sc vasp} package~\cite{vasp1,vasp2}. We have used the Perdew-Burke-Ernzerhof approximation to the exchange-correlation functional~\cite{PhysRevLett.77.3865}, an energy cut off of $400$\,eV and a $8\times8\times8$ $\mathbf{k}$-point grid to sample the electronic Brillouin zone. The CsPbCl$_3$ cubic lattice parameter of $5.605$\,\AA\@ is taken from Ref.~\cite{cspbcl3_1958_structure}. The phonon calculations use the finite displacement method~\cite{phonon_finite_displacement} in conjunction with nondiagonal supercells~\cite{non_diagonal}. The finite temperature band structure is calculated by performing Monte Carlo importance sampling over the harmonic vibrational density, and evaluating the band structure at each frozen-phonon configuration.

\begin{figure} \centering
\includegraphics[width=0.48\textwidth]{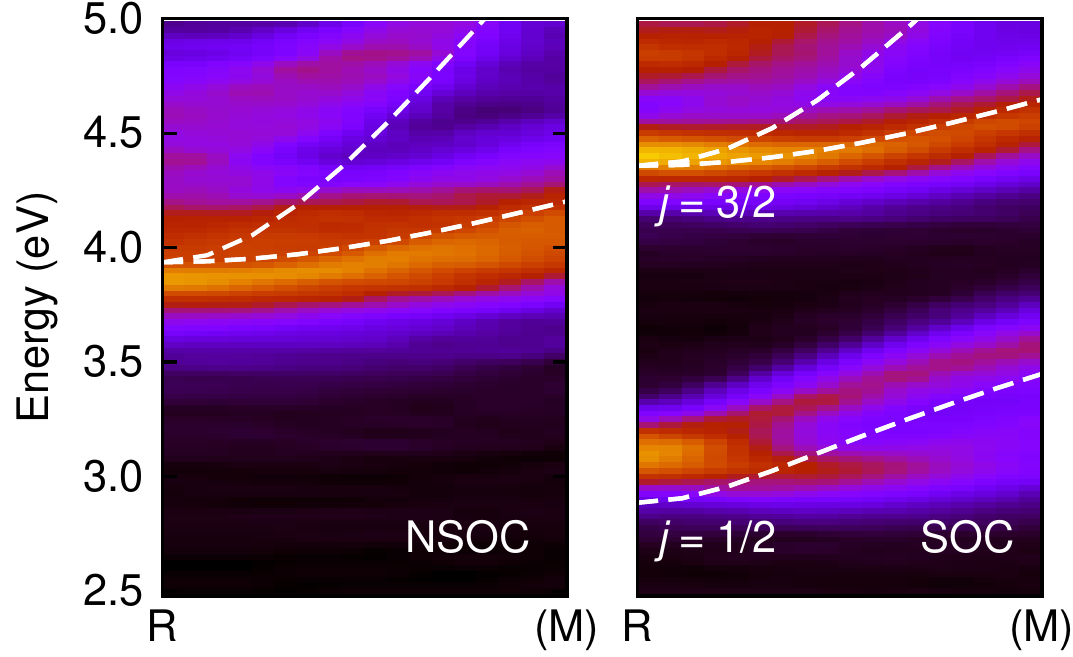}
\caption{Energy dispersion of the conduction bands of CsPbCl$_3$ assuming that the nuclei are static (white dashed lines) and calculating the thermal average in the harmonic approximation at $350$\,K (color gradient). The left diagram corresponds to a calculation neglecting the spin-orbit coupling (NSOC) and the right diagram includes the spin-orbit coupling (SOC). We have shifted the bands by $+2.0$\,eV to correct the standard DFT band-gap underestimation~\cite{experimental_gap_cspbcl3}.}
\label{fig:cspbcl3}
\end{figure}

CsPbCl$_3$ is a direct band gap semiconductor whose minimum
band gap occurs at the $R$ $(1/2,1/2,1/2)$
point, with the valence band dominated by hybridized Pb $s$ and Cl
$p$ states, and the conduction band dominated by Pb $p$ states. In
the absence of spin-orbit coupling, the six conduction-band $p_x$,
$p_y$, $p_z$ states, including spin, are degenerate, owing to the
fact that $R$ has the full cubic symmetry of the system. Moving away
from the $R$ point generally splits the six-fold degeneracy into
pairs of spin degenerate bands. Depending on the symmetry of the
ray in $\mathbf{k}$-space, a higher level of degeneracy
might be present. When nuclear vibrations are included the bands
become smeared about their static-lattice value, but maintain their
spin degeneracy. This behaviour is illustrated in
Fig.~\ref{fig:cspbcl3} (left pannel) showing the dispersion of the
conduction bands of CsPbCl$_3$ along the line from $R$ to $M$
$(1/2,1/2,0)$ (note that only a portion of this line is shown). The
static-lattice results are shown as dashed white lines, and the
color gradient depicts the results at $350$\,K.

When the spin-orbit interaction is included, the bands at $R$ split
into a $j=3/2$ quartet and a $j=1/2$ doublet, and the latter becomes
the conduction-band minimum. At the static-lattice level the system
is centrosymmetric. Moving away from $R$ in the Brillouin zone
splits the $j=3/2$ bands into pairs of bands with
$m_j=\pm3/2$ and $m_j=\pm1/2$, but these remain spin-degenerate, as do
the $j=1/2$ bands. This is shown
as the white dashed lines in Fig.~\ref{fig:cspbcl3} (right pannel).
When nuclear vibrations are included, the $j=1/2$ band is spin
split, with a behaviour analogous to that predicted with the
two-band model. In particular, the minimum of the conduction band is
shifted in both energy and momentum from the static-lattice minimum
at the $R$ point. We also observe that the band energy at $R$ is
blue-shifted at $350$\,K compared to the static-lattice counterpart.
This arises from the temperature dependence of the non-spin-orbit
coupling terms, which was neglected in the two-band model, but is
fully incorporated in the first principles calculations. The blue
shift, although opposite to what is reported in most semiconductors,
is consistent with that observed in other halide
perovskites~\cite{Foley_perovskites_2015,Milot_2015_temperature_mapbi3,patrick_cssni3_2015,mapbi3_monserrat}.

\begin{figure}
\centering
\includegraphics[width=0.48\textwidth]{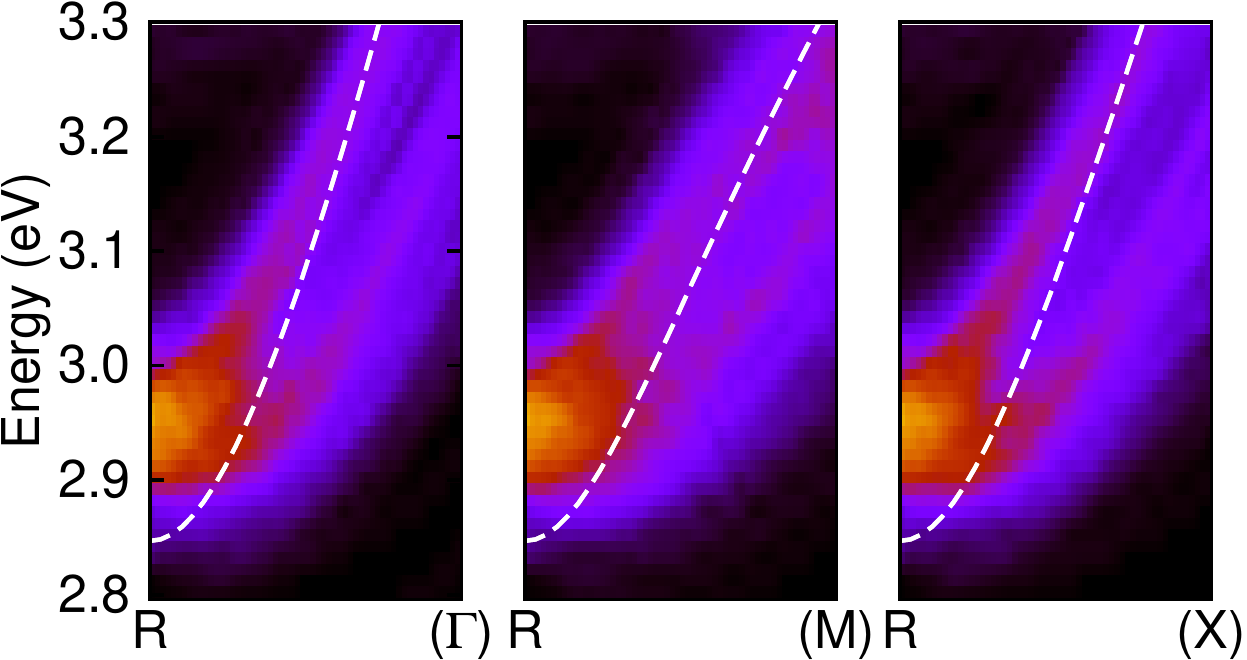}
\caption{Energy dispersion of the conduction bands of CsPbCl$_3$ assuming that the nuclei are static (white dashed lines) and calculating the thermal average in the harmonic approximation at $350$\,K (color gradient).
The dispersion is shown in the region near the band-gap minimum at the $R$ point along three high-symmetry directions in the vibrational BZ. The calculations have been performed using a $2\times2\times2$ supercell and we have shifted the bands by $+2.0$\,eV to correct the standard DFT band-gap underestimation~\cite{experimental_gap_cspbcl3}.}
\label{fig:2x2x2}
\end{figure}

The results reported in Fig.~\ref{fig:cspbcl3} only include the phonons at the center of the vibrational BZ. This is because the resulting spin splitting is clearer in this case, but spin splitting survives when a finer sampling is performed, as shown in Fig.~\ref{fig:2x2x2} for a $2\times2\times2$ vibrational BZ grid. The main difference arising from the inclusion of more vibrational modes is the reduction of the spin splitting magnitude, to a size of about $100$\,meV. Calculations using a grid of size $3\times3\times3$ show a similar spin splitting to that reported in Fig.~\ref{fig:2x2x2}. %We also note that spin splitting is observed along all directions in the electronic BZ, as shown in Fig.~\ref{fig:cspbcl3} along three high symmetry lines. 

CsPbCl$_3$ provides a clear demonstration of phonon-assisted spin splitting in centrosymmetric crystals. The predicted spin splitting is large, of the order of $100$\,meV, a feature that we attribute to the strong spin-orbit coupling exhibited by the lead halides~\cite{mapbbr3_rashba_exp}. We emphasize that this phenomenology is a general result, and all centrosymmetric crystals should exhibit phonon-induced spin splittings. However, as the magnitude of the splitting depends on the spin-orbit coupling strength, it might be impossible to observe this phenomenon in many materials. We also note that the closely related hybrid perovskites, which are being intensely investigated for photovoltaic applications~\cite{snaith_science_perovskite,perovskite_snaith_review}, are predicted to exhibit proper spin splitting due to the dipolar nature of the organic cation~\cite{sanvito_bs_mapbi3,rappe_soc_mapbi3,deangelis_soc_mapbi3,walsh_soc_mapbi3}. Our CsPbCl$_3$ results suggest that in these materials a spin-averaged spin splitting should be observed for timescales longer than the molecular rotation timescale. This is because the rotational motion of the dipolar cation averages over the dipole direction, and the resulting structure is centrosymmetric. 

We next propose a number of experiments to probe phonon-assisted spin splitting. The most direct measure would be provided by angle resolved photoelectron spectroscopy (ARPES). ARPES would be most appropriate for materials in which the phonon-assisted spin splitting is large in the valence bands. In the cesium lead halides CsPb$X_3$ ($X=$ Cl, Br, I) high-temperature cubic phase, it is the conduction $j=1/2$ Pb $p$ band that exhibits the strongest effect, so inverse ARPES would be the most appropriate probe for these materials.

Optical probes might also be useful to investigate phonon-assisted spin splitting. The $\mathbf{k}$-shift of the minimum band gap induced by spin splitting leads to a slightly indirect band gap in CsPbCl$_3$. As the static centrosymmetric structure of CsPbCl$_3$ is predicted to be a direct band gap material, observing indirect band gap behaviour in the optical absorption of the cesium lead halides would provide evidence for the predicted spin splitting. We nonetheless note that this observation might be preempted by the fact that the energy scale associated with the lattice vibrations is similar to the spin splitting energy scale (see Figs.~\ref{fig:cspbcl3} and \ref{fig:2x2x2}). A second optical feature could be provided by electron-doped samples with the Fermi level placed between the spin-split bands. Optical absorption between the spin-split bands should then be allowed, and this would appear as sub-gap absorption. An analogous signature in the optical conductivity of the bismuth tellurohalides has been used to study the inversion symmetry breaking Rashba splitting in these materials~\cite{bitei_optical_exp_theory}.

%Phonon-assisted spin splitting also expands the number of materials that could exhibit interesting spin-related phenomena in other areas. For example, the manipulation of spin is central in spintronics, and our work suggests that ultrafast probes could be used to access the spin degree of freedom even in time reversal and inversion symmetric materials.

In summary, we propose that spin splitting can occur even in time reversal and centrosymmetric materials with a combination of spin-orbit coupling and electron-phonon coupling. We have demonstrated this phenomenology using a simple two-band model, as well as realistic first principles calculations of CsPbCl$_3$. We expect that phonon-assisted spin splitting will expand the number of materials exhibiting interesting spin-related phenomena, in areas ranging from spintronics to photovoltaics.

\acknowledgments

This work was funded by NSF grant DMR-1408838.
B.M. acknowledges Robinson College, Cambridge, and the Cambridge
Philosophical Society for a Henslow Research Fellowship. 

\bibliography{dyn-spin}

\end{document}